\documentclass[twocolumn,showpacs,preprintnumbers,amsmath,amssymb]{revtex4}

\usepackage{graphicx}
\usepackage{epstopdf}
\usepackage{dcolumn}
\usepackage{bm}

\begin{document}

\preprint{}
\title{A Relativistic Flux-tube Model for  Hybrid Mesons}
\author{P. J. S. Watson}
\email{watson@physics.carleton.ca}
\affiliation{Ottawa-Carleton Institute for Physics, Carleton University, \\ Ottawa, Canada, K1S 5B6
}

\date{\today}

\begin{abstract}A number of authors have considered potential models for hybrid mesons. These frequently involve approximating the vibrating flux-tube by a set of beads, and making an adiabatic approximation which gives rise to a static inter-quark potential which has an effective repulsive 
$\frac{1}{r}$
 potential. We show that this approximation is almost certainly wrong. Since the beads are presumably massless, the correct approximation requires the solution of a Klein-Gordon-like equation treating  the beads and the quarks on the same footing. We show how to solve this in the one-bead case for massless quarks, and find the spectrum has an unexpected degeneracy. We generalise this to the $N_B$-bead case, which can still be solved exactly for massless quarks, and show how to renormalize the energy to obtain a plausible spectrum. We give a generic method to solve the equation for  massive quarks, and use this to derive a different non-relativistic equation. The results show a smooth behaviour both with respect to the quark mass and the number of beads.
  \end{abstract}
\pacs{}
\maketitle

\section{Introduction}
\begin{figure}[htbp] 
    \label{hybrid1}
  \includegraphics[width=2.5in]{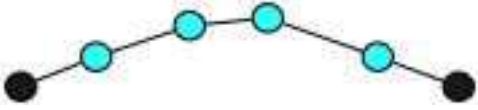} 
   \caption{A hybrid meson model}
\end{figure}

Many authors have considered models for hybrid mesons.  In most cases, the method used is similar in spirit to the original work of Isgur and Paton \cite{Isgur:1984bm}. In its simplest form, the meson is regarded as a 
$q\bar q$
 state with an interaction potential produced by an oscillating string or flux-tube. A physical model for this consists of massive beads linked by a linear potential (Fig \ref{hybrid1})
 These are treated non-relativistically in an adiabatic fashion and the oscillations are restricted to being perpendicular to the line joining the quarks. 
 	This basic idea has provided a rich source of models. There are justifications for various models from lattice-gauge theory (\cite{Bali:2000zv},\cite{Morningstar},\cite{Allen:1998wp}).
 More sophisticated analyses include non-adiabatic corrections  \cite{Merlin:1985mu}.  and a number of phenomenological analyses: see e.g.(\cite{Swanson:2003kg}, \cite{Close:2003ae}). The work by Barnes, Close and Swanson \cite{Barnes:1995hc} discusses a non-adiabatic model which has a similar starting point to the discussion below.

 This paper discusses a formulation which escapes from some of these restrictions. Firstly we show how to eliminate the centre of mass from a N-bead non-relativistic model. We use this to examine the adiabatic approximation and show that it can lead to highly misleading results in a very simple case. Then we write down a model where both the quarks and beads are treated as Klein-Gordon particles. The resulting equation can be exactly solved in the case where both quarks and beads are massless, and we write down a variational expression for the case where the quarks are massive. There are a number of subtleties in this model and the spectrum has an unexpected form. We have not made an attempt to turn this into a realistic model for conventional and hybrid mesons.
 
\section{Centre of Mass diagonalization}
The assumption that the quark motions are independent of the flux tube is not particularly logical for small quark mass.  We therefore consider the extension of the flux-tube hybrid model to include the quarks. If we assume the quarks and the beads have small masses, we can write
\[
\begin{array}{l}
 \mu _n  = \frac{{m_f }}{N},n = 2..N - 1,\mu _1  = m_1  + \frac{{m_f }}{N},\mu _N  = m_N  + \frac{{m_f }}{N} \\ 
 M = m_f  + m_1  + m_N  \\ 
 \end{array}
\]
where the flux tube mass 
${m_f}$ need not be a constant and we are allowing for the possibility of the quarks having different masses.

The full Schr\"odinger equations is
\[
\left( {\frac{1}{{2m_1 }}\nabla _1^2  + \frac{1}{{2m_N }}\nabla _N^2  + \sum\limits_{k = 2}^{N - 1} {\frac{1}{{2m_b }}\nabla _k^2 }  + \sigma l_N \left( {r,y_n } \right)} \right)\Psi  = E\Psi 
\]
In the spirit of the adiabatic approximation, we could assume that the beads are restricted to oscillations perpendicular to the tube, so ${\nabla _k^2}$ is 2-D, but ${\nabla _1^2}$ and ${\nabla _N^2}$ are 3 dimensional, so the quarks can move in 3-D. 

If the quarks are light, they need to be treated on the same footing as the beads, which means we  must separate the centre-of-mass motion from the relative motion terms in the sum over the KE. The CoM given by
\[
\vec R = \frac{{\mu \sum\limits_{n = 2}^{N - 1} {\vec r_i }  + m_1 \vec r_1  + m_N \vec r_N }}{M}
\]

 Writing
\begin{equation}
\label{eigen1}
\hat r_1  = R = \frac{{\sum\limits_{n = 1}^N {\mu _i \vec r_i } }}{M},\hat r_i  = \vec r_i  - \vec r_{i - 1} (i \ge 2)
\end{equation}
eliminates any cross terms of the form.
$
\frac{{\partial ^{} }}{{\partial R}}\frac{{\partial ^{} }}{{\partial \hat r_i }}
$
Note that this reduces to the usual expression for the 2-body case. This can be inverted, via
\[
\begin{array}{l}
 \vec r_N  = R + \frac{{\mu _1 \hat r_2 }}{M} + \frac{{\left( {\mu _1  + \mu _2 } \right)\hat r_3 }}{M} + .. + \frac{{\left( {\mu _1  + \mu _2  + ..\mu _{N - 1} } \right)\hat r_N }}{M} \\ 
 \vec r_{i - 1}  = \vec r_i  - \hat r_i  \\ 
 \end{array}
\]
This gives 
\[
\sum\limits_{n = 1}^N {\frac{1}{{2\mu _n }}\nabla _{r_n }^2 }  = \frac{1}{{2M}}\nabla _R^2  + \sum\limits_{n = 1}^N {\frac{1}{{2\mu _n }}\left( {\nabla _{\hat r_n }^{}  - \nabla _{\hat r_{n + 1} }^{} } \right)} ^2 
\]

We now want to diagonalize the KE operator. Write 
\[\label{KE_G}
\sum\limits_{n = 1}^N {\frac{1}{{2\mu _n }}\left( {\nabla _{\hat r_n }^{}  - \nabla _{\hat r_{n + 1} }^{} } \right)} ^2  = \partial _{\bf{i}} ^{\bf{T}} {\bf{G}}\partial _{\bf{i}} 
\]
where the matrix G is symmetric and tri-diagonal.

\begin{equation}
\label{GG}
{\bf{G}} = \left[ {\begin{array}{*{20}c}
   {\frac{1}{{\mu _1 }} + \frac{1}{{\mu _2 }}} & { - \frac{1}{{\mu _2 }}} & {} & {}  \\
   { - \frac{1}{{\mu _2 }}} & {\frac{1}{{\mu _2 }} + \frac{1}{{\mu _3 }}} & { - \frac{1}{{\mu _3 }}} & {}  \\
   {} & { - \frac{1}{{\mu _3 }}} & {\frac{1}{{\mu _3 }} + \frac{1}{{\mu _4 }}} & {..}  \\
   {} & {} & {..} & {..}  \\
\end{array}} \right]
\end{equation}

We can then diagonalize this using 
\[
{\bf{\hat u}} = {\bf{S\hat r}},{\bf{S}}^T {\bf{GS}} = \Lambda 
\]
which allows  the kinetic energy to be written
\[
\frac{1}{{2m_R }}\frac{{\partial ^2 }}{{\partial r^2 }} + \sum\limits_{n = 1}^N {\frac{1}{{2\hat \mu _n }}\nabla _{u_n }^2 } 
\]
It is then possible to solve the equation analytically if we make two further assumptions: that the motion is adiabatic, so that the beads oscillate perpendicular to the line joining the quarks and that  P.E. 
\[
l_N \left( {r,\hat y_n } \right) = \sigma \sum\limits_{i = 1}^N {\left| {\hat r_i } \right|}  = \sigma \sum\limits_{i = 1}^N {\left[ {\hat y_i^2  + \left( {\frac{r}{N}} \right)^2 } \right]} ^{1/2} 
\]
(where ${\hat y_i }$ is perpendicular to the line joining the quarks)
can be expanded: i.e.
\[
V_N \left( {r,\hat r_n } \right) \approx r + \frac{N}{{2r}}\sum\limits_{i = 1}^N {\hat y_i^2 } 
\]
so the flux tube interaction gives rise to an oscillator, but with a strength that increases with N. 

Since 
${\bf{\hat u}} = {\bf{S\hat r}}$
, the P.E. term can be written 
$\sum\limits_{i = 1}^N {\left[ {\hat r_i^2 } \right]}  = \sum\limits_{i = 1}^N {\left[ {\hat u_i^2 } \right]} $
This allows the N-bead equations to be separated into N uncoupled equations, where the effective spring constant is given by 
\[
k_{eff}^2  = \frac{{\sigma N}}{{\bar r}}
\]

\section{Failure of the adiabatic method: Zero-bead case}
We would like  to check the validity of this expansion in a very simple example: the 0 bead case which be solved exactly in both 3-D and 2+1 D methods. The basic equation is 
\begin{equation}
\label{3-D}
\left( {\frac{{\hbar ^2 }}{{2m_R }}\nabla ^2  - \sigma r + E_{nl} } \right)\psi _{nl} \left( {\vec r} \right) = 0
\end{equation}

\subsubsection{Exact solution}
As is well known, for l = 0, after conversion to a dimensionless form via
\[
y = \beta \left( {z - z_n } \right) , z_n  = \frac{{y_n }}{\beta }  ,  \beta  = \left( {\frac{{2m\sigma }}{{\hbar ^2 }}} \right)^{1/3} 
\]

one obtains the  Airy equation giving 
\[
E_n  = \left( {\frac{{\hbar ^2 \sigma ^2 }}{{2m}}} \right)^{1/3} y_n 
\]
 where  $y_n$ is the n-th zero of the Airy function
\subsubsection{Conventional Variational Solution}
We can, of course, solve (\ref{3-D}) by a standard variational approach with a Gaussian-wave function. The variational energy 
\[
E = \frac{{\hbar ^2 }}{{2m}}\beta ^2 \frac{3}{2} + \frac{{2\sigma }}{{\beta \sqrt \pi  }}
\]
gives 
\[
\beta  = {\rm{0}}{\rm{.867}}\left( {\frac{{\sigma m}}{{\hbar ^2 }}} \right)^{1/3} ,E = 1.865\left( {\frac{{\hbar ^2 \sigma ^2 }}{m}} \right)^{1/3} 
\]
\subsubsection{2+1-D solution:}
We can now solve the 3-D equation by a ``phonon'' approx. This means put 
$\vec r = \left( {x,\hat y} \right)$
 so 
$$\left| {\vec r} \right| = \sqrt {x^2  + \hat y^2 }  \approx x + \frac{{\hat y^2 }}{{2x}}$$
and the trial wave function is
$$\phi _0\left( {x,\vec r} \right)=e^{-\left( {\beta ''x} \right)^2}e^{-\left( {\beta '\left( x \right)r} \right)^2}$$

The 2-D ``phonon'' equation has the form
\[
\left( {\frac{{\hbar ^2 }}{{2m_R }}\left( {\frac{{\partial ^2 }}{{\partial y^2 }} + \frac{1}{y}\frac{\partial }{{\partial r}} - \frac{{\ell ^2 }}{{y^2 }}} \right) - \frac{{\sigma y^2 }}{{2x}} + E_{nl} } \right)\psi _{nl} \left( r \right) = 0
\]
with the ground state solution
\[
\beta ' = \frac{1}{\hbar }\sqrt {\frac{{\sigma m}}{x}} ,E'\left( x \right) = \hbar \sqrt {\frac{\sigma }{{mx}}} 
\]
This then gives the corresponding 1-D equation
\[
\left( {\frac{{\hbar ^2 }}{{2m}}\frac{{\partial ^2 }}{{\partial x^2 }} - \sigma x + \hbar \sqrt {\frac{\sigma }{{mx}}}  + E} \right)\phi \left( x \right) = 0
\]
with a variational solution 
\[
E = \frac{{\hbar ^2 }}{{2m}}\frac{{\beta ^2 }}{2} + \frac{{g_V \sigma }}{{\sqrt \pi  \beta }} + \frac{{\Gamma \left( {\frac{1}{4}} \right)}}{{\sqrt \pi  }}\left( {\frac{\sigma }{m}} \right)^{1/2} \beta ^{1/2} 
\]
which can be minimised numerically for $\beta$ and E.

\subsubsection{Simultaneous minimisation}
	 To justify a technique we will use later, we can minimise over both $\beta ,\beta '$ simultaneously. This means doing a non-linear minimisation for both the phonon and longitudinal modes: i.e. split the equation into transverse (phonon) and longitudinal motion as before 
$$\begin{array}{l}
 \left( { - \frac{{\hbar ^2 }}{{2m}}\left( {\frac{{d^2 }}{{dx^2 }} + \frac{{\partial ^2 }}{{\partial y^2 }} + \frac{1}{{y^2 }}\frac{{\partial ^2 }}{{\partial \phi ^2 }}} \right) + \frac{1}{2}\sigma \sqrt {x^2  + \hat y^2 } } \right)\phi _0 \left( {x,\hat y} \right) \\ 
  = E\phi _0 \left( {x,\hat y} \right) \\ 
 \end{array}$$
and using a trial wave-function
$\phi _0 \left( {x,\hat y} \right) = e^{ - \left( {\beta x} \right)^2 /2} e^{ - \left( {\beta 'y} \right)^2 /2} $
, perform simultaneous non-linear minimisation on $\beta ,\beta '$
\subsubsection{Results}

The results, using $m_q= 300 MeV,\sigma =180000 MeV^2\ $, are shown in table \ref{results}. 
\begin{table}[htdp]
 \caption{Comparison of zero-bead results}
\begin{center}
 \begin{tabular}{@{} lll @{}}
    \hline
    ¥ & $\beta$ & Energy \\ 
    \hline
    3-D exact  & 1.3203 & 1113.45\\ 
    3-D variational  & 1.3850 & 1116.63\\ 
    2+1-D phonon approximation & 0.89668 & 1567.87 \\ 
    2+1-D minimization & 1.3846 & 1116.62 \\ 
    \hline
  \end{tabular}
\end{center}
\label{results}
 \end{table}%
We conclude that the phonon approximation is disastrously bad. It can be argued that the model is unrealistic, since the transverse motion could not be regarded as adiabatic in this case. However, we find similar results for a one-bead example, but there are no exact results to compare it.
\section{Klein-Gordon Equation}
The assumption that the quarks can be treated non-relativistically is not sensible when the quark mass is small. Further, it is hard to justify using a  non-relativistic approximation in the limit that $N\to \infty $ since the bead mass goes to zero and the Schr\"odinger equation becomes meaningless. Hence we would like to have an equation which works for both massless beads and massless quarks. 

We consider a Klein-Gordon equation for the both 
\begin{equation}
\label{KG5}
\left( {p_i^2 c^2  + m_i^2 c^4 } \right)\psi  = 0,m_i  = \mu _i  + \frac{{V_i }}{{c^2 }}
\end{equation}

where we have assumed that the confining potential transforms as a Lorentz scalar (assuming it transforms as the 4th component of a 4-vector leads to other problems: see \cite{ram:549}). Obviously a realistic model would require a Dirac equation for the quarks. 

The spectrum should be well-behaved and  independent of the mass in the $\mu_0\to 0$ limit, since the K-G equation becomes 
\begin{equation}
\label{KG_0}
\left( { - \hbar ^2 \nabla ^2  - \frac{{E^2 }}{{c^2 }} + V^2 } \right)\psi  = 0
\end{equation}

We need to reduce the n-body K-G equation to a Schr\"odinger-like equation with no relative time. This does not appear to be in the literature. One cannot use a conventional reduced mass reduction: see (\cite{Ebert:1997nk}).

If we have n particles with individual momenta $p_i$, we can obviously write the total momentum
$$P_\mu =\sum {p_{i\mu }}$$
which is correct relativistically or non-relativistically. Each bead or quark satisfies
\[
\left( {{\bf{p}}_i ^2 c^2  + m_i ^2 c^4 } \right)\psi  = \varepsilon _i^2 \psi ,m_i  = \mu  + \frac{{V_i }}{{c^2 }}
\]
 so by analogy with the preceding we write 
\[
p_i  = \frac{{\varepsilon_i }}{E}P + \hat q_i  - \hat q_{i - 1} 
\]
Here and in what follows we use E to refer to total system energies and $\varepsilon $ to refer to individual component (quark, bead or phonon) energies. This is specified more carefully in Appendix A.

 Then the ${\hat q}$'s can be reconstructed
\[
\begin{array}{l}
 \hat q_1  = p_1  - \frac{{\varepsilon _1 }}{E}P, \\ 
 \hat q_k  = p_k  + \hat q_{k - 1}  - \frac{{\varepsilon _k }}{E}P = \sum\limits_{}^k {p_i }  - \frac{{\sum\limits_{}^k {\varepsilon _i } }}{E}P \\ 
 \end{array}
\]

 Note that  $\hat q_k $ is still purely space-like:  and the total energy 
$
E = \sum\limits_{i = 1}^n {\varepsilon _i } 
$
. Effectively in this process we are replacing the masses of the non-relativistic particles with the corresponding energies.

Hence 

\begin{equation}
\label{KG-n}
\sum\limits_{}^{} {\frac{{p_i ^2 }}{{E_i }}}  = \frac{{P^2 }}{E} + \sum\limits_{}^{} {\frac{{\left( {\hat q_i  - \hat q_{i - 1} } \right)^2 }}{{E_i }}} 
\end{equation}

so the overall K-G equation becomes

\begin{equation}
\label{KG_3}
\begin{array}{l}
 \sum\limits_{}^{} {\frac{1}{{\varepsilon _i }}\left( {p_i ^2 c^2  + m_i ^2 c^4 } \right)\Psi }  \\ 
  = \left( {\frac{{P^2 c^2 }}{E} + \sum\limits_{}^{} {\frac{{\left( {\hat q_i  - \hat q_{i - 1} } \right)^2 c^2  + \left( {\mu _i c^2  + V_i } \right)^2 }}{{\varepsilon _i }}} } \right)\Psi  = E\Psi  \\ 
 \end{array}
\end{equation}

This equation is correct but apparently useless, since the individual 
${\varepsilon _i }$
 are unknown: in the equivalent NR treatment, the energies are replaced by masses whaich are, of course, known \textit{a priori}. Below we show how to find the ${\varepsilon _i }$ by a variational trick.
  
 \section{Massless quarks}
 We consider in turn the (trivial) zero-bead solution, the  one bead solution and show how the solution can be generalised to the $N_b$-bead solution. Then we show how the $N_b$-bead solution can be renormalized to give a sensible spectrum. The starting assumption is that there is a scalar confining  potential, $V = \sigma r$
equally divided between the quarks and the beads, so that (\ref{KG5}) becomes
\[
m_i ^2 c^4  = \mu _i^2 c^4  + \mu _i^{} \sigma rc^2  + \frac{{\sigma ^2 r^2 }}{4} \Rightarrow \frac{{\sigma ^2 r^2 }}{4}
\] 
\subsection{Massless zero-bead solution}
The energy equations simplify to 
\[
E = Mc^2 ,E_R  = \frac{{Mc^2 }}{2},\mu _R  = \frac{M}{2},\varepsilon _1  = \varepsilon _2  = \frac{{Mc^2 }}{2}
\]
Then K-G eqn becomes
\[
\left( { - \frac{{4\hbar ^2 c^2 }}{E}\nabla ^2  + \frac{{\sigma ^2 r^2 }}{E}} \right)\psi  = E\psi 
\]
 with an immediate solution 
\[
\beta ^2  = \frac{\sigma }{{2\hbar c}},E^2  = 6\hbar c\sigma 
\]
 
\subsection{Massless One-Bead}
\begin{figure}[htbp] 
  \includegraphics[width=2.5in]{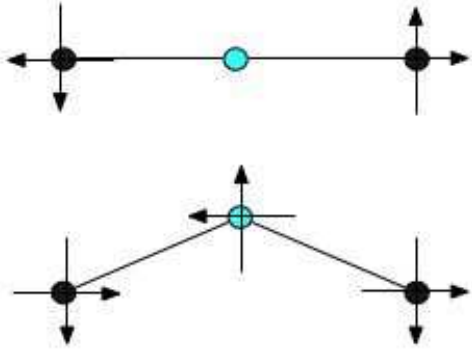} 
  \caption{Two modes for one-bead model}
   \label{modes}
\end{figure}

The interparticle potential becomes 
$
V_k \left( {\hat r_n } \right) = \frac{\sigma }{2}\left( {\hat r_k^{}  + \hat r_{k - 1}^{} } \right)
$. Hence the sum over potentials is 
\[
\begin{array}{l}
 V^2  = \frac{{\sigma ^2 }}{4}\left( {\frac{{\hat r_1^2 }}{{\varepsilon _1 }} + \frac{{\hat r_1^2  + 2\hat r_2^{} \hat r_1^{}  + \hat r_2^2 }}{{\varepsilon _2 }} + \frac{{\hat r_2^2 }}{{\varepsilon _3 }}} \right) \\ 
  = \frac{{\sigma ^2 }}{4}\left( {\left( {\hat r_1^2  + \hat r_2^2 } \right)\left( {\frac{1}{{\varepsilon _1 }} + \frac{1}{{\varepsilon _2 }}} \right) + \frac{{2\hat r_2^{} \hat r_1^{} }}{{\varepsilon _2 }}} \right) \\ 
 \end{array}
\]
giving
\[
\left( { - \hbar ^2 c^2 \sum\limits_i^{} {\frac{{\left( {\nabla _i  - \nabla _{i - 1} } \right)^2 }}{{\varepsilon_i }}}  + V^2 } \right)\psi  = Mc^2 \psi 
\]
 The K.E. can be diagonalized with eigenvalues 
$1/\varepsilon_ +  ,1/\varepsilon_ -  $
given by 
\[
\begin{array}{l}
 E = 2\varepsilon _1  + \varepsilon _2 ,\varepsilon _ +   = \varepsilon _1 ,\varepsilon _ -   = \frac{{\varepsilon _1 \varepsilon _2 }}{{2\varepsilon _1  + \varepsilon _2 }} = \frac{{\varepsilon _1 \varepsilon _2 }}{E}, \\ 
 \frac{1}{{\varepsilon _1 }} + \frac{1}{{\varepsilon _2 }} = \frac{1}{2}\left( {\frac{1}{{\varepsilon _ +  }} + \frac{1}{{\varepsilon _ -  }}} \right) \\ 
 \hat u_ \pm   = \frac{1}{{\sqrt 2 }}\left( {\hat r_1  \pm \hat r_2 } \right),\hat r_1^2  + \hat r_2^2  = \hat u_ + ^2  + \hat u_ - ^2  \\ 
 2\hat r_2^{} \hat r_1^{}  = \left( {\hat u_ + ^2  - \hat u_ - ^2 } \right) \\ 
 \end{array}
\]
(note that this is just the relativistic analog of the result in  Merlin and Paton \cite{Merlin:1985mu}. Hence the K-G equation becomes
\[
\left( { - \hbar ^2 c^2 \left( {\frac{{\nabla _ +  ^2 }}{{\varepsilon _ +  }} + \frac{{\nabla _ -  ^2 }}{{\varepsilon _ -  }}} \right) + \frac{{\sigma ^2 }}{4}\left[ {\hat u_ +  ^2 \frac{1}{{\varepsilon _ -  }} + \hat u_ -  ^2 \frac{1}{{\varepsilon _ +  }}} \right]} \right)\psi  = E\psi 
\]
which separates into two phonon equations:
\[
\left( { - \frac{{\hbar ^2 c^2 }}{{\varepsilon _ \pm  }}\nabla _ \pm  ^2  + \frac{{\sigma ^2 }}{{4\varepsilon _ \mp  }}\hat u_ \pm  ^2 } \right)\psi  = \hat \varepsilon _ \pm  \psi 
\]
Note the distinction between the $\varepsilon_i$ (individual energies of the quarks and bead) and $ \hat  \varepsilon_ \pm$ (individual energies of the phonon states). This has obvious solutions:
\[
\begin{array}{l}
 \beta _ \pm  ^2  = \frac{\sigma }{{2c\hbar }}\left( {\frac{{\varepsilon _ \pm  }}{{\varepsilon _ \mp  }}} \right)^{1/2} ,\hat \varepsilon _ \pm   = \frac{{c\hbar \sigma }}{{2\sqrt {\varepsilon _ +  \varepsilon _ -  } }}\left( {2n_ \pm   + l_ \div   + \frac{3}{2}} \right), \\ 
 E_0  = \hat \varepsilon _ +   + \hat \varepsilon _ -   = \varepsilon _ +   + \varepsilon _ -   = \frac{{3c\hbar \sigma }}{{4\sqrt {\varepsilon _ +  \varepsilon _ -  } }} \\ 
 \end{array}
\]
To solve this, we will assume that the quarks carry fractions $\alpha _1$ of the total energy of the system, and regard  $\alpha _1$ as a variational parameter, so
\[
\begin{array}{l}
 \varepsilon_1  = \alpha _1 E = \varepsilon_3  = \varepsilon_ +  ,\varepsilon_2  = \alpha _B E,\alpha _B  = (1 - 2\alpha _1 ) \\ 
 \varepsilon_ -   = \alpha _1 \alpha _B E \Rightarrow E^2 \left( {\alpha _1 } \right) = \frac{{3c\hbar \sigma }}{{\sqrt {\alpha _1 ^2 (1 - 2\alpha _1 )} }} \\ 
 \end{array}
\]
gives a minimum value of $\alpha _1 = 1/3$: i.e. the quarks and the bead carry an equal fraction of the energy, which is what we expect. This gives 
\[
\beta _ +  ^2  = \frac{\sigma }{{2c\hbar }}\left( {\frac{1}{3}} \right)^{1/2} ,\beta _ -  ^2  = \frac{\sigma }{{2c\hbar }}\left( 3 \right)^{1/2} ,E^2  = \frac{3}{2}\sqrt {27} c\hbar \sigma 
\]
Physically we can interpret the two solutions as show in fig \ref{modes}: the + solution corresponds to the two quarks moving in opposite direction while the bead remains stationary at the centre of mass, while the - solution corresponds to the quarks moving in the same direction but opposite to the bead. 

It is worth noting the peculiar feature of the solution: the two phonon states have the same excitation spectrum but different $\beta$'s. 

\subsection{Massless n-bead solution}
The overall K-G equation (\ref{KG_3}) becomes
\[
\left( {\sum\limits_{}^{} {\frac{{\left( {\hat q_i  - \hat q_{i - 1} } \right)^2 c^2 }}{{E_i }}}  + \frac{{\sigma ^2 }}{4}\sum\limits_{}^{} {\frac{{\left( {\hat r_i  + \hat r_{i - 1} } \right)^2 }}{{E_i }}} } \right)\Psi  = E\Psi 
\]

\textbf{If} we assume that the ${\varepsilon_i }$'s are equal 
\begin{equation}
\label{def}
\varepsilon_i  = \frac{E}{N+1}
\end{equation}
 then this can be solved (we justify the assumption below). The previous method generalises: 
\[
\begin{array}{l}
 \left( { - \frac{{\hbar ^2 c^2 N}}{E}\sum\limits_{}^{} {\left( {\nabla _i  - \nabla _{i - 1} } \right)^2 }  + \frac{N}{E}\frac{{\sigma ^2 }}{4}\sum\limits_{}^{} {\left( {\hat r_i  + \hat r_{i - 1} } \right)^2 } } \right)\psi  =  \\ 
 \left( { - \frac{{\hbar ^2 c^2 N}}{E}\sum\limits_{}^{} {\nabla _i G_{ij} \nabla _j }  + \frac{N}{E}\frac{{\sigma ^2 }}{4}\sum\limits_{}^{} {\hat r_i G'_{ij} \hat r_j } } \right)\psi  = E\psi  \\ 
 \end{array}
\]
where
\[
G = \left[ {\begin{array}{*{20}c}
   2 & { - 1} & {} & {}  \\
   { - 1} & 2 & { - 1} & {}  \\
   {} & { - 1} & 2 & { - 1}  \\
   {} & {} & { - 1} & 2  \\
\end{array}} \right],G' = \left[ {\begin{array}{*{20}c}
   2 & 1 & {} & {}  \\
   1 & 2 & 1 & {}  \\
   {} & 1 & 2 & 1  \\
   {} & {} & 1 & 2  \\
\end{array}} \right]
\]
and 
\begin{equation}
\label{eigen2}
\hat u = {\bf{S}}\hat r,{\bf{S}}^T {\bf{GS}} = \Lambda
\end{equation}
The structure is obviously similar to (\ref{GG}).

	The KE term can be diagonalized  as before. The same operation that digaonalizes G also diagonalizes $G'$. The eigenvalues are given by \cite{Yueh}, \footnote{This result was discovered ``experimentally''. It can be shown by the algorithms developed by Yueh (\cite{Yueh})}. 
\[
\Lambda _N^i  = \left( {2\sin \left( {\frac{{\pi i}}{{2\left( {N + 1} \right)}}} \right)} \right)^2 
\]

This then separates to give a set of N uncoupled ``phonon'' equations of the form
\[
\left( { - \frac{{\hbar ^2 c^2 N}}{E}\Lambda _i^N \nabla _{u_i }^2  + \frac{N}{E}\frac{{\sigma ^2 }}{4}\Lambda _{N - i}^N \hat u_i^2 } \right)\psi  = \hat \varepsilon_i \psi 
\]
which have the usual SHO form, with different interpretations of the energy and potential. With a modification of the usual substitutions

\begin{equation}
\label{NR_1}
\begin{array}{l}
 \left( {\beta _i ^N } \right)^2  = \sqrt {\frac{{\Lambda _{N - i} }}{{\Lambda _i }}} \frac{\sigma }{{2\hbar }} = \cot \left( {\frac{{\pi i}}{{N + 1}}} \right)\frac{\sigma }{{2\hbar }} \\ 
 \hat \varepsilon ^N _i  = \frac{{c\hbar \sigma \left( {N + 1} \right)}}{{2E^N }}\left( {2n_i  + l_i  + \frac{3}{2}} \right)\sqrt {\Lambda _i \Lambda _{N - i} }  \\ 
  = \frac{{c\hbar \sigma \left( {N + 1} \right)}}{{2E^N }}\left( {2n_i  + l_i  + \frac{3}{2}} \right)\sin \left( {\frac{{\pi i}}{{N + 1}}} \right) \\ 
 \end{array}
\end{equation}

we can write the solution as
\[
\psi \left( {\hat r_i } \right) = A\zeta _i ^l L_{n_i }^l \left( {\zeta _i ^2 } \right)e^{ - \zeta _i ^2 } ,\zeta _i  = \beta _i \hat u_i 
\]

The energies can be summed to give  the total energy  
\[
\begin{array}{l}
 E^N  = \sum\limits_{i = 1}^N {\hat \varepsilon _i^N }  = \frac{3}{2}\frac{{c\hbar \sigma \left( {N + 1} \right)}}{{2E^N }}\sum\limits_{i = 1}^N {\sin \left( {\frac{{\pi i}}{{N + 1}}} \right) = }  \\ 
  = \frac{3}{2}\frac{{c\hbar \sigma \left( {N + 1} \right)}}{{2E^N }}\frac{{\sin \left( {\frac{\pi }{{N + 1}}} \right)}}{{1 - \cos \left( {\frac{\pi }{{N + 1}}} \right)}} \\ 
 \end{array}
\]
 so 
\[
\left( {E^N } \right)^2  \approx \frac{3}{2}\frac{{c\hbar \sigma }}{{\pi E^N }}\left( {N + 1} \right)^2 
\]

There are several problems with this solution. Not surprisingly, the total energy is divergent as $N \to \infty $. Secondly the curious degeneracy noted above remains: the energy of the $n^{th}$ and $(N-n)^{th}$ levels are equal, even though they are described by different values of $\beta$. This violates one's intuitive idea that large  $\beta$ corresponds to small E and vice-versa. Finally, the individual energies of the phonon modes are
\[
\hat \varepsilon^N _i  \approx \sqrt {\frac{{3c\hbar \sigma }}{2}} \frac{{\pi ^{1/2} }}{2}\sin \left( {\frac{{\pi i}}{{N + 1}}} \right) \approx \sqrt {\frac{{3c\hbar \sigma }}{2}} \frac{{\pi ^{3/2} i}}{{2\left( {N + 1} \right)}}
\]
so the excitation energies are vanishingly small in the $N \to \infty $ limit. 

\subsection{Energy Renormalization}
In order to obtain finite answers, we subtract a energy $\Delta E^N $ from the total non-renormalized energy $E^N $ given by (\ref{NR_1})
to give a (fixed) renormalized energy $\tilde E^0$ for the ground state: 
\[
\tilde E^0  = E^N  - \Delta E^N 
\]

By writing  
\[
\Delta E^N  = \sum\limits_{i = 1}^N {\hat \varepsilon_i^N \left( {0,0} \right)}  - \tilde E^0 
\]
gives the renormalized energy of each phonon mode as 

\begin{equation}
\label{ren_E_2}
\tilde \varepsilon^N _i \left( {n_i ,l_i } \right) = \frac{{c\hbar \sigma \left( {N + 1} \right)}}{{2\tilde E^0 }}\left( {2n_i  + l_i } \right)\sin \left( {\frac{{\pi i}}{{N + 1}}} \right)
\end{equation}

This now has a sensible limit for all modes in the $N \to \infty$ limit
\[
\tilde \varepsilon ^\infty  _i \left( {n_i ,l_i } \right) = \frac{{c\hbar \sigma }}{{2\tilde E^0 }}\left( {2n_i  + l_i } \right)\pi i
\]
It makes sense to have the energy fixed by the zero-bead solution, so

\begin{equation}
\label{E_0}
\tilde E^0  = \sqrt {6\hbar c\sigma } 
\end{equation}

and the total energy of the excited phonon states is then given (in the $N \to \infty $ limit) by
\begin{equation}
\label{E_ren}
E^\infty  \left( {n_k ,l_k } \right) = \sqrt {6c\hbar \sigma } \left( {1 + \frac{\pi }{{12}}\sum\limits_{}^{} {\left( {2n_i  + l_i } \right)i} } \right)
\end{equation}

\section{Massive quarks}
We have ignored the quark masses: this introduces two  extra types of term, via 
\[
m_i ^2 c^4  = \mu _i^2 c^4  + \mu _i^{} \sigma rc^2  + \frac{{\sigma ^2 r^2 }}{4}
\]
The second term, in particular, causes considerable trouble since it mixes the phonon states.  We provide a very rapid numerical algorithm for taking this into account. In principle one can include a bead-mass: there is no need to do so, and we show that one can in fact find a non-relativistic reduction of the equation even with a massless bead. Again, we analyse the 0-bead, 1-bead and n-bead solutions. 
\subsection{Massive quarks, zero-bead case}
We consider the general unequal-mass case. The K-G equation is \begin{widetext}
\[
\left( { - \frac{{\hbar ^2 c^2 }}{{\varepsilon_R }}\nabla ^2  + \frac{1}{{\varepsilon_1 }}\left( {\mu _1 ^2 c^4  + \mu _1 c^2 \sigma r + \frac{1}{4}\sigma ^2 r^2 } \right) + \frac{1}{{\varepsilon_2 }}\left( {\mu _2 ^2 c^4  + \mu _2 c^2 \sigma r + \frac{1}{4}\sigma ^2 r^2 } \right)} \right)\psi  = E\psi 
\]
\end{widetext}
In order to solve all of this variationally (and to connect this easy example with later equations), we proceed as follows. 
\begin{enumerate}
\item Assume that the quarks carry fractions $\alpha _1$, $\alpha _2$ of the total energy of the system. Then 
\[
\varepsilon_1  = \alpha _1 E,\varepsilon_2  = \alpha _2 E,\varepsilon_R  = \alpha _1 \alpha _2 E,\alpha _1  + \alpha _2  = 1
\]
 obviously 
$0 < \alpha _1 ,\alpha _2  < 1$
\item Hence the K-G equation becomes
\begin{equation}
\label{KG-0}
\begin{array}{l}
 \left( { - \frac{{\hbar ^2 c^2 }}{{\alpha _1 \alpha _2 }}\nabla ^2  + \frac{{\mu _1 ^2 c^4 }}{{\alpha _1 }} + \frac{{\mu _2 ^2 c^4 }}{{\alpha _2 }} + c^2 \sigma r\left( {\frac{{\mu _1 }}{{\alpha _1 }} + \frac{{\mu _2 }}{{\alpha _2 }}} \right) + \frac{{\sigma ^2 r^2 }}{{4\alpha _1 \alpha _2 }}} \right)\psi  \\ 
  = E^2 \psi  \\ 
 \end{array}
\end{equation}
\item Use a Gaussian trial wave-function, as suggested by the massless case,  giving the variational energy: 
\begin{equation}
\label{KG-1}
\begin{array}{l}
 E^2  = \frac{{\hbar ^2 c^2 }}{{\alpha _1 \alpha _2 }}\frac{3}{2}\beta ^2  + \frac{{\mu _1 ^2 c^4 }}{{\alpha _1 }} + \frac{{\mu _2 ^2 c^4 }}{{\alpha _2 }} +  \\ 
 c^2 \sigma \left( {\frac{{\mu _1 }}{{\alpha _1 }} + \frac{{\mu _2 }}{{\alpha _2 }}} \right)\frac{2}{{\beta \sqrt \pi  }} + \frac{{\sigma ^2 }}{4}\frac{3}{{2\alpha _1 \alpha _2 \beta ^2 }} \\ 
 \end{array}
\end{equation}

\end{enumerate}
This obviously contains the massless case. 
\subsubsection{Non-relativistic Reduction}
It is not obvious that this technique gives a sensible non-relativistic reduction. If the masses are large and equal, 
$
M = 2\mu  + \frac{{E_B }}{{c^2 }}
$ where $E_B$ is the binding energy. Putting 
\[
\begin{array}{l}
 E = Mc^2  = \mu _1 c^2  + \mu _2 c^2  + E_B , \\ 
 \varepsilon _1  \approx \mu _1 c^2 ,\varepsilon _2  \approx \mu _2 c^2 ,\varepsilon _R  \approx \frac{{\mu _1 \mu _2 c^2 }}{{\mu _1  + \mu _2 }} = \mu _R c^2  \\ 
 \end{array}
\]
Ignoring terms with 
$V^2,E_B^2$ 
 in  (\ref{KG-0}) gives 
\[
\left( { - \frac{{\hbar ^2 }}{{2\mu _R }}\nabla ^2  + \sigma r - E_B } \right)\psi  = 0
\]
 which is the Schrodinger equation.

However, we wish to use a variational method, and it is not conventional to find the  reduced mass  by this technique! We get a mass term from (\ref{KG-1}) of the form 
$\alpha _1 \alpha _2 \left( {\frac{{\mu _1 ^2 c^4 }}{{\alpha _1 }} + \frac{{\mu _2 ^2 c^4 }}{{\alpha _2 }} - E^2 } \right)$.
 In the infinite mass limit, this gives an apparent mass term
\[
\begin{array}{l}
 \hat m^2  = \left( {\mu _1 ^2 \left( {1 - \alpha _1 } \right)^2  + \mu _2 ^2 \alpha _1 ^2  - 2\mu _1 \left( {1 - \alpha _1 } \right)\mu _2 \alpha _1 } \right) \\ 
  = \left( {\mu _1 \left( {1 - \alpha _1 } \right) - \mu _2 \alpha _1 } \right)^2  \\ 
 \end{array}
\]
 Minimising this gives (obviously) 
$\alpha _1  = \frac{{\mu _1 }}{{\mu _1  + \mu _2 }}$
so the ``reduced energy'' becomes 
\[
\varepsilon _R  = \alpha _1 \alpha _2 E = \mu _R c^2 
\]
as expected.
\subsubsection{Coulomb-interaction and non-zero angular momentum}
We can generalise this  by including a Coulomb-type interaction (see Olsson \cite{Olsson:1997fm})  
$V(r) = \frac{\kappa }{r}
$ which is the fourth component of a 4-vector, so that 
\[
\varepsilon_1 ^2  \to \left( {\varepsilon_1  - \frac{\kappa }{{2r}}} \right)^2  = \varepsilon_1 \left( {\varepsilon_1  - \frac{\kappa }{r} + \frac{{\kappa ^2 }}{{\varepsilon_1 4r^2 }}} \right)
\]
This adds a term 
\[
 - \frac{{2\kappa E}}{r} + \frac{{\kappa ^2 }}{{4\alpha _1 \alpha _2 r^2 }}
\]
to the K-G equation. 

We can also generalise this to any angular momentum $\ell$: both these generalisations require us to use a modified trial wave function of the form 
\[
\psi \left( r \right) = Cr^p e^{ - \frac{{\beta ^2 r^2 }}{2}} 
\]
 We will write the variational energy in the generic form 
\[
\begin{array}{l}
 E^2 \left( {\mu _1 ,\mu _2 ,\ell } \right) = \hbar ^2 c^2 d_{ - 2} \left( {\alpha _i } \right)\beta ^2  + \kappa Ed_{ - 1} \left( \beta  \right) +  \\ 
 c^4 d_0 \left( {\mu _j ,\alpha _i } \right) + \sigma c^2 d_1 \left( \beta  \right) + \frac{{\sigma ^2 }}{4}\frac{{d_2 ^k \left( {\alpha _i } \right)}}{{\beta ^2 }} \\ 
 \end{array}
\]
where 
\begin{equation}
\label{d_s}
\begin{array}{l}
 d_{ - 2}  = \frac{1}{{\left( {p + \frac{1}{2}} \right)}}\left( {\frac{{4l\left( {l + 1} \right) + 4p + 3}}{{2\alpha _1 \alpha _2 }} + \frac{{\kappa ^2 }}{4}} \right) \\ 
 d_{ - 1}  = 2\beta \frac{{\Gamma \left( {p + 1} \right)}}{{\Gamma \left( {p + \frac{3}{2}} \right)}} \\ 
 d_0  = \frac{{\mu _1 ^2 }}{{\alpha _1 }} + \frac{{\mu _2 ^2 }}{{\alpha _2 }} \\ 
 d_1  = \left( {\frac{{\mu _1 }}{{\alpha _1 }} + \frac{{\mu _1 }}{{\alpha _2 }}} \right)\frac{{\Gamma \left( {p + 2} \right)}}{{\beta \Gamma \left( {p + \frac{3}{2}} \right)}} \\ 
 d_2  = \frac{1}{4}\left( {p + \frac{3}{2}} \right) \\ 
 \end{array}
\end{equation}

This generic form is useful since it generalises to the one-bead and n-bead case, and it can be solved by a rapid algorithm which is outlined in appendix B. For the purposes of renormalizing the n-bead energy later, we regard the zero-bead energy as the renormalized energy of the ground state
\[
\tilde E^0  = E\left( {\mu _1 ,\mu _2 ,\ell } \right)
\]
 in the sense of (\ref{E_0}). 
\subsection{Massive quarks: one-bead }
To solve the unequal-mass one-bead case, we will assume that the quarks carry fractions $\alpha _1$, $\alpha _3$ of the total energy of the system. 
\[
\varepsilon_1  = \alpha _1 E,\varepsilon_2  = \alpha _B E,\varepsilon_3  = \alpha _3 E,\alpha _B  = (1 - \alpha _1  - \alpha _3 )
\]
This introduces two extra terms. There is a  constant term with same 

\begin{equation}
\label{M_1}
d_0  = \frac{{\mu _1 ^2 c^4 }}{{\alpha _1 }} + \frac{{\mu _2 ^2 c^4 }}{{\alpha _2 }}
\end{equation}
 as in (\ref{d_s}). The second term is

\begin{equation}
\label{M_2}
2c^2 \left( {\frac{{\mu _1 V_1 }}{{\varepsilon_1 }} + \frac{{\mu _N V_N }}{{\varepsilon_3 }}} \right) = c^2 \frac{\sigma }{E}\left( {\frac{{\mu _1 \left| {\hat r_1 } \right|}}{{\alpha _1 }} + \frac{{\mu _N \left| {\hat r_2 } \right|}}{{\alpha _3 }}} \right)
\end{equation}

is more difficult to handle

In  the equal mass case, (\ref{M_1}) becomes: 
$$d_0  = 2\frac{{\mu _1 ^2 }}{{\alpha }}$$ and the linear term (\ref{M_2}) becomes
\[
\sqrt 2 \frac{{\mu \sigma }}{{\alpha _1 E}}\left| {\hat u_ + ^2  + \hat u_ - ^2 } \right|^{1/2} \left( {1 + O\left( {\cos ^2 \left( \theta  \right)} \right) + ...} \right)
\]

Ignoring the higher order terms gives  
\[
\label{M_3}
\frac{1}{{\alpha _1 E}}\left( \begin{array}{l}
  - \hbar ^2 c^2 \left( {\nabla _ +  ^2  + \frac{{\nabla _ -  ^2 }}{{\alpha _B }}} \right) + 2\mu ^2 c^4  +  \\ 
 \sqrt 2 c^2 \mu \sigma \left| {\hat u_ + ^2  + \hat u_ - ^2 } \right|^{1/2}  + \frac{{\sigma ^2 }}{4}\left[ {\frac{{\hat u_ +  ^2 }}{{\alpha _B }} + \hat u_ -  ^2 } \right] \\ 
 \end{array} \right)\psi  = E\psi 
\]
The linear term (\ref{M_2}) gives us a new term in the variational energy 
\[
\begin{array}{l}
 A^2 \sqrt 2 c^2 \frac{{\mu \sigma }}{{E_1 }}\int {\int {e^{ - \beta _ +  ^2 u_ + ^2 } e^{ - \beta _ -  ^2 u_ - ^2 } \left| {\hat u_ + ^2  + \hat u_ - ^2 } \right|^{1/2} u_ + ^2 u_ - ^2 du_ + ^{} du_ - ^{} } }  \\ 
  = \sqrt 2 c^2 \frac{{\mu \sigma }}{{\alpha _1 E}}f_{00}^1 \left( {\beta _ +  ,\beta _ -  } \right) \\ 
 \end{array}
\]
which can be written as 
$d_1 \left( {\alpha ,\beta } \right) = \sqrt 2 c^2 \frac{{\mu \sigma }}{{\alpha E}}f_{00}^1 \left( {\beta _ +  ,\beta _ -  } \right)$
in the notation of (\ref{d_s}). We show how to evaluate $f_{00}^1$ in Appendix C. . Hence the overall variational solution becomes
\[
\begin{array}{l}
 E^2  = \frac{3}{2}\frac{{\hbar ^2 c^2 }}{{\alpha _1 }}\left( {\beta _ +  ^2  + \frac{{\beta _ -  ^2 }}{{\alpha _1 \alpha _B }}} \right) + 2\mu ^2 c^4  +  \\ 
 \sqrt 2 c^2 \frac{{\mu \sigma }}{{\alpha _1 }}f_{00}^1 \left( {\beta _ +  ,\beta _ -  } \right) + \frac{{3\sigma ^2 }}{{8\alpha _1 }}\left( {\frac{1}{{\alpha _B \beta _ +  ^2 }} + \frac{1}{{\beta _ -  ^2 }}} \right) \\ 
 \end{array}
\]
 which can be rapidly solved by the technique of Appendix A. 
 \subsubsection{Non-relativistic reduction}
It is useful to have a non-relativistic reduction of this equation. The conventional reduction of allowing the bead and quark masses to become large will not work, since we believe the bead mass is zero..  We need to reduce 
(\ref{M_3}) to a corresponding NR equation for comparison. Putting $E = 2\mu c^2  + E_B $ gives\[
\begin{array}{l}
 \alpha _1 E^2  - 2\mu ^2 c^4  \approx 2\mu ^2 c^4 \left( {2\alpha _1  - 1} \right) + 4\alpha _1 \mu c^2 E_B  \\ 
  = 4\alpha _1 \mu c^2 E_B  - 2\mu ^2 c^4 \alpha _B  \\ 
 \end{array}
\]
The K-G equation then becomes
\[
\begin{array}{l}
 \frac{1}{{\alpha _1 }}\left( { - \frac{{\hbar ^2 }}{{4\mu }}\left( {\nabla _ +  ^2  + \frac{{\nabla _ -  ^2 }}{{\alpha _B }}} \right) + \frac{1}{2}\mu c^2 \alpha _B  + \frac{1}{{2\sqrt 2 }}\sigma \left| {\hat u_ + ^2  + \hat u_ - ^2 } \right|^{1/2} } \right)\psi  \\ 
  = E_B \psi  \\ 
 \end{array}
\]
so the bead has acquired a pseudo-mass $\alpha _1 \alpha _B \mu $
 
This has the variational solution
\[
\begin{array}{l}
 E_B  = \frac{3}{2}\frac{{\hbar ^2 }}{{4\mu }}\left( {\frac{{\beta _ +  ^2 }}{{\alpha _1 }} + \frac{{\beta _ -  ^2 }}{{\alpha _B \alpha _1 }}} \right) +  \\ 
 \frac{1}{{2\sqrt 2 \alpha _1 }}\sigma f_{00}^1 \left( {\beta _ +  ,\beta _ -  } \right) + \frac{1}{2}\mu c^2 \frac{{\alpha _B }}{{\alpha _1 }} \\ 
 \end{array}
\]
Note that the masslessness of the bead does not give rise to a Schr\"odinger equation with zero mass. The usual limit of $\alpha = 1/2$ which would minimise the mass term does not minimise the energy.

\subsection{Massive Quarks: n-bead case}
We want $N_B = N-2$ beads and 2 quarks, with different masses, so that we have N particles in all and the quarks have a fraction 
$\alpha _1 ,\alpha _N $
of the total energy, so each bead has a fraction

\begin{equation}
\label{alp}
\alpha _B  = \frac{{1 - \alpha _1  - \alpha _2 }}{{N - 2}}
\end{equation}
of the total. Now the K-G eqn becomes
\[
\sum\limits_{}^{} {\frac{{ - \hbar ^2 \left( {\nabla _i  - \nabla _{i - 1} } \right)^2 c^2  + \left( {\mu _i c^2  + V_i } \right)^2 }}{{E_i }}} \Psi  = E\Psi 
\]
The KE and  $V^2$ terms give expressions similar to (\ref{GG}) with G and $G'$ are now given by
\[
\begin{array}{l}
 G = \left[ {\begin{array}{*{20}c}
   {\alpha _1 ' + 1} & { - 1} & {} & {} & {}  \\
   { - 1} & 2 & { - 1} & {} & {}  \\
   {} & { - 1} & 2 & { - 1} & {...}  \\
   {} & {...} & {...} & {..} & {..}  \\
   {} & {} & {} & { - 1} & {\alpha _2 ' + 1}  \\
\end{array}} \right] \\ 
 G' = \left[ {\begin{array}{*{20}c}
   {\alpha _1 ' + 1} & 1 & {} & {} & {}  \\
   1 & 2 & 1 & {} & {}  \\
   {} & 1 & 2 & 1 & {...}  \\
   {} & {...} & {...} & {..} & {..}  \\
   {} & {} & {} & 1 & {\alpha _2 ' + 1}  \\
\end{array}} \right] \\ 
 \end{array}
\]
 with 
\[
\alpha _1 ' = \frac{{1 - \alpha _1  - \alpha _N }}{{\alpha _1 \left( {N - 2} \right)}} = \frac{{\alpha _B }}{{\alpha _1 }}
\]

The term in $\mu V$ becomes 
\[
2c^2 \left( {\frac{{\mu _1 V_1 }}{{\varepsilon _1 }} + \frac{{\mu _N V_N }}{{\varepsilon _N }}} \right) = \frac{{\sigma c^2 }}{E}\left( {\frac{{\mu _1 \left| {\hat r_1 } \right|}}{{\alpha _1 }} + \frac{{\mu _N \left| {\hat r_N } \right|}}{{\alpha _N }}} \right)
\]
Finally the mass term gives 
\[
c^4 \left( {\frac{{\mu _1 ^2 }}{{\varepsilon _1 }} + \frac{{\mu _N ^2 }}{{\varepsilon _N }}} \right) = \frac{{c^4 }}{E}\left( {\frac{{\mu _1 ^2 }}{{\alpha _1 }} + \frac{{\mu _1 ^2 }}{{\alpha _N }}} \right)
\]
 as before. 
There are several new difficulties: 
\begin{enumerate}
\item for non-equal masses, diagonalizing the KE matrix does not diagonalize the PE matrix.
\item the linear terms like $
{\mu _1 \left| {\hat r_1 } \right|}
$  involve all the $\hat u_i$'s, so the $\mu V$ is very complex.
\item there are $(N_b + 1)$ different  $\beta$'s (one  for each phonon mode) which must be found variationally.
\item in addition, we have $\alpha _1$ and $\alpha _N$ which must be regarded as variational parameters.
\end{enumerate}
Surprisingly, this can be approximately solved by a quite simple numerical technique: the algorithm is as follows.
\begin{enumerate}
\item Guess $\alpha _1, \alpha _N$
\item Diagonalize G and find eigenvectors $e_i^j$ \item Ignore off-diagonal terms in 
${\bf{e}}^{\bf{T}} {\bf{G'e}} = \Lambda '$
\item Write 
$
\hat r_1  = \sum\limits_{}^{} {e_j^{\left( 1 \right)} u_j } 
$
so that 
\[
\left| {\hat r_1 } \right| = \sqrt {\hat r_1 ^2 }  = \sqrt {\sum\limits_{}^{} {e_j^{\left( 1 \right)} e_k^{\left( 1 \right)} u_j u_k } } 
\]
\item Then (approximately) the expectation values required for the linear interaction can be found (See Appendix D):
\[
f_1 \left( {\beta _i } \right) = \left\langle {\left| {\hat r_1 } \right|} \right\rangle  \approx \left( {\sum\limits_{i = 1}^N {\left( {l_i  + \frac{3}{2}} \right)\left( {\frac{{e_i^{\left( 1 \right)} }}{{\beta _i }}} \right)^2 } } \right)^{1/2} 
\]

\item This gives us a total (non-renormalized) energy of the form:
\[
\begin{array}{l}
 E^2  = \hbar ^2 c^2 \sum\limits_j^{} {\Lambda _j \beta _j ^2 \left( {l_j  + \frac{3}{2}} \right)}  + \frac{{\mu _1 ^2 c^4 }}{{\alpha _1 }} + \frac{{\mu _1 ^2 c^4 }}{{\alpha _N }} \\ 
 2\sigma c^2 \left( {\frac{{\mu _1 }}{{\alpha _1 }}f_1 \left( {\beta _i } \right) + \frac{{\mu _N }}{{\alpha _N }}f_N \left( {\beta _i } \right)} \right) + \frac{{\sigma ^2 }}{4}\sum\limits_{}^{} {\frac{{\Lambda _j '}}{{\beta _j ^2 }}\left( {l_j  + \frac{3}{2}} \right)}  \\ 
 \end{array}
\]
to be minimised to find the 
$\beta _j $. Obviously if $\mu_1 = \mu_N = 0$ this separates: it still has the form of  \ref{var_E}.

\item After finding the energy $\beta _j $. the energy must still be minimised by varying the $\alpha_i$.  
\end{enumerate}
Although linear term mixes the phonon modes, it is  useful to define the non-renormalized energy of the phonon modes as
\begin{widetext}
\[
\hat \varepsilon _j  = \frac{1}{{\tilde E^0 }}\left( {\hbar ^2 c^2 \Lambda _j \beta _j ^2 \left( {l_j  + \frac{3}{2}} \right) + \frac{{\sigma ^2 }}{4}\frac{{\Lambda _j '}}{{\beta _j ^2 }}\left( {l_j  + \frac{3}{2}} \right) + \frac{1}{N}\left[ {2\sigma c^2 \left( {\frac{{\mu _1 }}{{\alpha _1 }}f_1 \left( {\beta _i } \right) + \frac{{\mu _N }}{{\alpha _N }}f_N \left( {\beta _i } \right)} \right) + \frac{{\mu _1 ^2 c^4 }}{{\alpha _1 }} + \frac{{\mu _2 ^2 c^4 }}{{\alpha _2 }}} \right]} \right)
\]
\end{widetext}
 where we have somewhat arbitrarily assigned 
$\frac{1}{N}$
 of the linear and mass energies to each mode. 

By setting the mass to zero, we can confirm the division of energies implied by (\ref{def}): i.e. the numerical minimisation does in fact divide the energies equally between the massless quarks and the beads.

\section{Results}
In this section we show some results arising from this model. The intention is \textbf{not} to give a detailed comparison with experiment (which would require inclusion of colour  Coulomb and magnetic interactions) but rather to show how the results change as we increase the number of beads and move away from the exactly soluble massless case. 

\begin{figure}[htbp] 
  \includegraphics[width=2in]{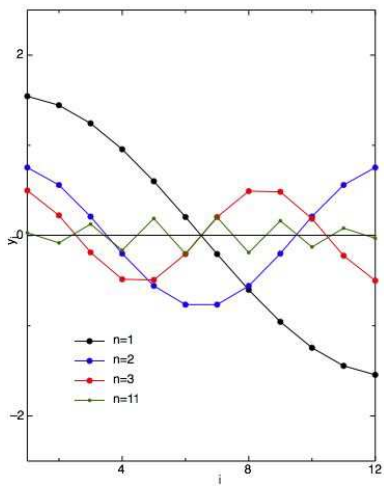} 
   \caption{Eigenmodes for Massless quarks }
   \label{eigenLL}
\end{figure}

\begin{figure}[htbp] 
  \includegraphics[width=2in]{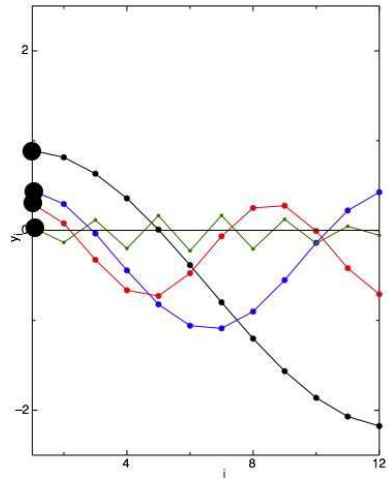} 
  \caption{Eigenmodes for one massive, one massless quark}
   \label{eigenHL}
\end{figure}

\begin{figure}[htbp] 
  \includegraphics[width=2in]{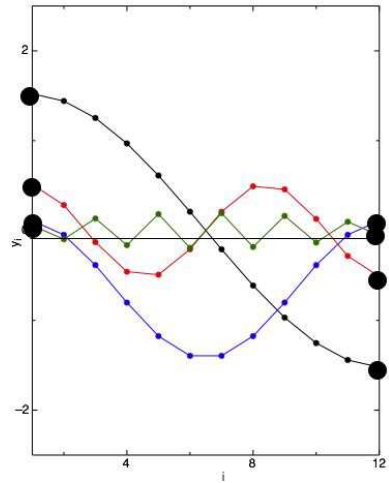} 
   \caption{Eigenmodes for two massive quarks}
   \label{eigenHH}
\end{figure}

It is possible to get an intuitive feeling for the phonon states by using (\ref{eigen2}) to reconstruct the eigenmodes. We show the results for the 10-bead case for 3 combinations of quark masses; the case with two massless quarks (fig (\ref{eigenLL})),  with one massless and one heavy quark and 2 equal mass heavy quarks  (fig (\ref{eigenHH})).  (fig (\ref{eigenHL}))

Note that the heavy quarks stay relatively fixed, while the massless  quarks move as much as the beads, as one would expect. The corresponding non-relativistic calculation shows simply a straight line for the  ground state solution for the case with two massive quarks: in the relativistic case the beads necessarily carry a considerable amount of the energy.

\begin{figure}[htbp] 
  \includegraphics[width=2.5in]{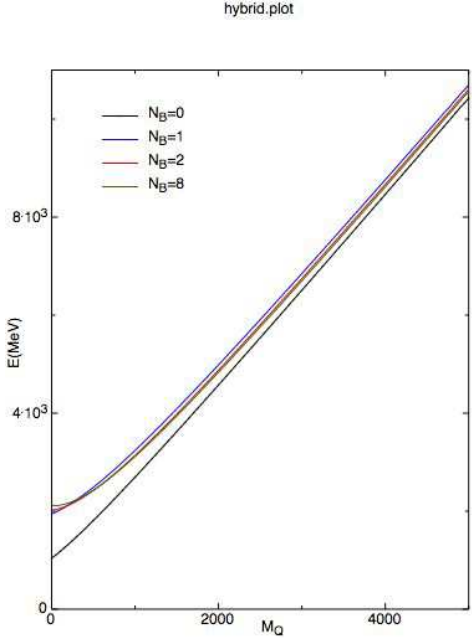} 
   \caption{Total energies for states consisting of two massive quarks and a  varying number of beads. The lowest curve is the ground state (identical for all $N_B$ by the renormalization process) and the upper curves show the energy for the $l_1 = 1$}
      \label{en1}
\end{figure}

The renomalised energies of the bound state are given by 
\[
\tilde E = \sum\limits_i^{} {\tilde \varepsilon ^N _i \left( {n_i ,l_i } \right)} 
\] where 
$
{\tilde \varepsilon ^N _i \left( {n_i ,l_i } \right)}
$
 is given by (\ref{NR_1}) in the massless case. The only excited states we  consider are the $l_1 = 1, l_{i>1}=0$ state, which corresponds to a conventional orbitally excited meson, and the $l_1 = 0,l_2 = 1, l_{i>2}=0$, which corresponds to the first hybrid state. If the quarks are massless, the ground state lies at 
$\tilde E^0  = \sqrt {6\hbar c\sigma }  \simeq 1GeV$
( this would be lowered in a realistic model by the short-range colour interactions). Fig (\ref{en1}) shows the energies of the  $l_1 = 1, l_{i>1}=0$ bound state  as a function of quark mass for equal mass quarks and a varying number of beads. Note that the mass is dominated by the quark mass.
 
\begin{figure}[htbp] 
  \includegraphics[width=2.5in]{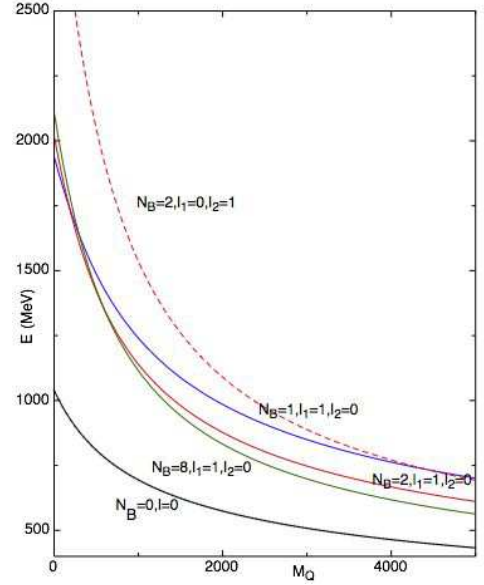} 
   \caption{Excitation energies for states consisting of 2 massive quarks and a  varying number of beads}
   \label{BE_HH}
\end{figure}

\begin{figure}[htbp] 
  \includegraphics[width=2.5in]{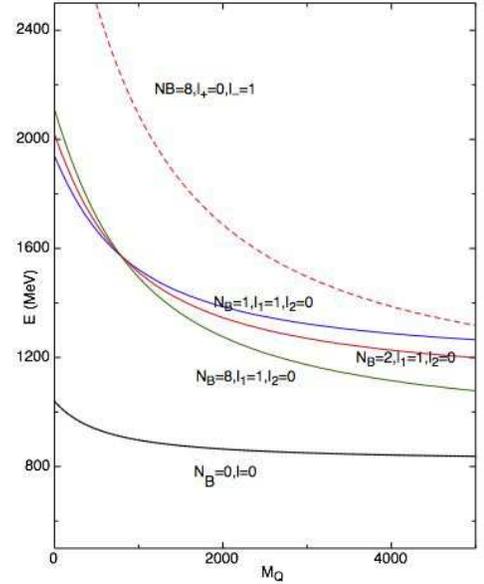} 
   \caption{Excitation energies for states consisting of one massive and one massless quark and a  varying number of beads}
      \label{BE_HL}
\end{figure}

 It is more sensible to plot the excitation energy (i.e. the total energy with the quark mass subtracted), and in Fig (\ref{BE_HH}) we show the ground state energy and the excitation energy of the 2 states for equal-mass quarks and varying number of beads.  
Fig (\ref{BE_HL}) is the corresponding plot for one heavy quark and one massless quark.  
The energies are roughly independent of the number of beads. The energies typically lie lower as the number of  beads increases. The hybrid state lies consistently at twice the excitation energy of the first orbitally excited state

\section{Conclusions}
We have derived a relativistic model for mesons in a bead approximation to a  flux-tube model. The model is analytically soluble for massless quarks, and can be renormalized to give a sensible spectrum. The massive quark case can be solved approximately  by a rapidly converging algorithm.
There are, however, several unattractive features. 
\begin{enumerate}
\item The algorithm based on (\ref{alp}) implies that the amount of energy carried by the quarks decreases as the number of beads increases. This is artificial, but it is not clear whether it has any real physical consequence.
\item It is somewhat difficult to include short-range corrections: the colour Coulomb can be incorporated by a comparatively minor variation, as indicated, but the (crucial) hyperfine interactions probably require the extension of (\ref{KG_0}) to include a Dirac equation for the quarks.
\item It is difficult to consider radial excitations in the massive quark case.
\item The choice of the renormalized ground state energy is somewhat arbitrary.
\end{enumerate}

These (and other) problems will be addressed in a later paper.

\begin{acknowledgments} 
This work was begun in collaboration with Drs. Frank Close and Jo Dudek, and I am very grateful to them for many conversations.
\end{acknowledgments}

\appendix
\section{Notation}
There are several different energies in the calculation. They can be enumerated as follows:
\begin{enumerate}
\item  $\varepsilon _i^{} $ is the energy of the $i^{th}$ object (quark or bead)
\item  $\hat \varepsilon _i^N \left( {n_i ,l_i } \right)$ is the non-renormalized energy of the $i^{th}$ phonon mode, excited to an  $\left( {n_i ,l_i } \right)$ state, in the N-particle approximation.
\item $\tilde \varepsilon _i^N \left( {n_i ,l_i } \right)$ is the renormalized energy  of the $i^{th}$ phonon mode in the N-particle approximation.
\item $E^N \left( {n_i ,l_i } \right) = \sum\limits_{i = 1}^N {\hat \varepsilon _i^N \left( {n_i ,l_i } \right)} $
is the non-renormalized energy  of the bound state.
\item $E^0  = E^N \left( {0,0} \right)$ is the ground state energy for the meson.
\item $\tilde E^N \left( {n_k ,l_k } \right) = \sum\limits_{i = 1}^N {\tilde \varepsilon _i^N \left( {n_i ,l_i } \right)} $ is the renormalized excited-state energy: i.e. the mass of the meson.
\end{enumerate}

\section{Generic Variational Solution}
In all the calculations, the variational value of the energy takes the following form: 
\begin{equation}
\label{var_E}
\begin{array}{l}
 E^2  = \sum\limits_j^{} {d_{ - 2} ^j \left( {\alpha _i } \right)\beta _j ^2 }  + \kappa Ed_{ - 1} \left( {\beta _k } \right) +  \\ 
 d_0 \left( {\mu _j ,\alpha _i } \right) + \sigma d_1 \left( {\beta _k ,\mu _j ,\alpha _i } \right) + \frac{{\sigma ^2 }}{4}\sum\limits_{}^{} {\frac{{d_2 ^k \left( {\alpha _i } \right)}}{{\beta _k ^2 }}}  \\ 
 \end{array}
\end{equation}

with the following interpretations:
\begin{enumerate}
\item 
${\beta _k }$
 are the wave-function parameters.: i.e. 
$\psi \left( {u_i } \right) = u_i ^{l_i } e^{ - \left( {\beta _i u_i } \right)^2 } $
\item The ${\alpha _i }$ are the energy separation parameters: see (\ref{alp}
\item ${d_{ - 2} ^k \left( {\alpha _i } \right)}$ is the KE term: e.g for the 2-bead case with no Coulomb interaction,
\[
\begin{array}{l}
 d_{ - 2} ^ +  \left( {\alpha _i } \right) = \frac{{\hbar ^2 c^2 }}{{\alpha _1 \alpha _B }}\left( {\frac{3}{2} + l_ +  } \right) \\ 
 d_{ - 2} ^ -  \left( {\alpha _i } \right) = \frac{{\hbar ^2 c^2 }}{{\alpha _B }}\left( {\frac{3}{2} + l_ -  } \right) \\ 
 \end{array}
\]
\item ${d_{ - 1} \left( {\beta _k } \right)}$ is Coulomb correction: note that it contains E since it is the 4th component of a 4-vector.
\item ${d_0 \left( {\mu _j ,\alpha _i } \right)}$ is the mass term:
\[
d_0 \left( {\mu _j ,\alpha _i } \right) = \frac{{\mu _1 ^2 }}{{\alpha _1 }} + \frac{{\mu _2 ^2 }}{{\alpha _2 }}
\]
 so vanishes for massless quarks, 
\item 
${d_1 \left( {\alpha _i ,\mu _j ,\beta _k } \right)}$
is the linear potential term: 
\[
d_1 \left( {\alpha _i ,\mu _j ,\beta _k } \right) = \left( {\frac{{\mu _1 }}{{\alpha _1 }} + \frac{{\mu _2 }}{{\alpha _2 }}} \right)\frac{2}{{\beta \sqrt \pi  }}
\]

for 0-bead and 
\[
\begin{array}{l}
 d_1 \left( {\alpha _i ,\mu _j ,\beta _i } \right) = \left( {\frac{{\mu _1 }}{{\alpha _1 }}f_{}^{\left( 1 \right)}  + \frac{{\mu _N }}{{\alpha _N }}f_{}^{\left( N \right)} } \right), \\ 
 f_{}^{\left( m \right)}  \approx \sqrt {\frac{3}{2}} \left( {\sum\limits_i^N {\frac{{e_{im}^2 }}{{\beta _i ^2 }}} } \right)^{1/2}  \\ 
 \end{array}
\]
for the $N_B$-bead case.
\item 
${d_2 ^k \left( {\alpha _i } \right)}$ arises from square of the linear terms: 
\[
d_2 ^ +  \left( {\alpha _i } \right) = \left( {\frac{3}{2} + l_ -  } \right),d_2 ^ -  \left( {\alpha _i } \right) = \frac{{\left( {\frac{3}{2} + l_ +  } \right)}}{{\alpha _B }}
\]
for 1-bead case
\end{enumerate}
The merit of writing it in this form is that it is completely general and has an immediate solution for massless quarks with no Coulomb interaction:
\[
d_{ - 2} ^i \left( {\alpha _i } \right) = \Lambda _i^N ,d_2 ^i \left( {\alpha _i } \right) = \Lambda _{N - i}^N ,d_{ - 1} ^i  = d_0 ^i  = d_1 ^i  = 0
\]
 gives 
\[
\left( {\beta _i ^N } \right)^2  = \sqrt {\frac{{\Lambda _{N - i} }}{{\Lambda _i }}} \frac{\sigma }{2}
\]
The general solution for the non-linear equation set can be written 
\begin{equation}
\label{beta_k}
\begin{array}{l}
 \beta _k  = \left[ {\frac{{\frac{{2\sigma ^2 }}{4}d_2 ^k \left( {\alpha _i } \right) - \kappa E\left( {\alpha _i ,\beta _k } \right)\beta _k ^3 \frac{{\partial d_{ - 1} \left( {\beta _i } \right)}}{{\partial \beta _k }} + \sigma \beta _k ^3 \frac{{\partial d_1 \left( {\beta _i } \right)}}{{\partial \beta _k }}}}{{2d_{ - 2} ^k \left( {\alpha _i } \right)}}} \right]^{1/4}  \\ 
  = H_k \left( {\beta _i ,\beta _k } \right) \\ 
 \end{array}
\end{equation}
This is the basis of the algorithm for rapid solution of the general case: \begin{enumerate}
\item Start with the above 
\[
\beta _{i0} ^{}  = \left( {\frac{{d_2 ^i \left( {\alpha _i } \right)}}{{d_{ - 2} ^i \left( {\alpha _i } \right)}}} \right)^{1/4} 
\]
\item Fix all ${\beta _i }$ except for one, ${\beta _k }$.
\item Find 3 successive values of ${\beta _k }$ using eqn. (\ref{beta_k}) 
\[
\beta _k ^{\left( n \right)}  = H\left( {\beta _i ,\beta _k ^{\left( {n - 1} \right)} } \right)
\]
\item FInd the Aiken extrapolation to provide an updated ${\beta _k }$ 
\[
\beta _k  = \frac{{\beta _k ^{\left( 2 \right)} \beta _k ^{\left( 0 \right)}  - \left( {\beta _k ^{\left( 1 \right)} } \right)^2 }}{{\beta _k ^{\left( 2 \right)}  - 2\beta _k ^{\left( 1 \right)}  + \beta _k ^{\left( 0 \right)} }}
\]
\item Choose another ${\beta _{k'} }$ and iterate until the values converge (typically 3 iterations for the full set).
\item Use the final values of  the ${\beta _i  }$ to find the energy E using \ref{var_E}.
\end{enumerate}
Ths works very rapidly because 
\[
\frac{{\beta _k ^3 \partial d_1 \left( {\alpha _i ,\mu _j ,\beta _i } \right)}}{{\partial \beta _k }} =  - \frac{3}{2}\left( {\frac{{\mu _1 }}{{\alpha _1 }}\frac{{e_{i1}^2 }}{{f_{}^{\left( 1 \right)} }} + \frac{{\mu _N }}{{\alpha _N }}\frac{{e_{iN}^2 }}{{f_{}^{\left( N \right)} }}} \right)
\]
is a comparatively slowly varying function of the ${\beta _i  }$. 

There is an extra complication when the Coulomb term is included because the energy needs to be calculated after each of the ${\beta _i  }$ are estimated. Note that this format of the equation retains the phonon picture.

\section{Standard Integrals}
For the one-bead case, we need to define a standard integral
\[
\begin{array}{l}
 g_{l_ +  l_ -  }^k \left( {\beta _ +  ,\beta _ -  } \right) =  \\ 
 \int {\int {u_ + ^{2l_ +  } e^{ - \beta _ +  ^2 u_ + ^2 } u_ - ^{2l_ -  } e^{ - \beta _ -  ^2 u_ - ^2 } \left| {\hat u_ + ^2  + \hat u_ - ^2 } \right|^{k/2} u_ + ^2 u_ - ^2 du_ + ^{} du_ - ^{} } }  \\ 
  = \int {\int \begin{array}{l}
 e^{ - y^2 \left( {\beta _ + ^2 \cos ^2 \left( \phi  \right) + \beta _ - ^2 \sin ^2 \left( \phi  \right)} \right)}  \\ 
 \cos ^{2 + 2l_ +  } \left( \phi  \right)\sin ^{2 + 2l_ -  } \left( \phi  \right)y^{5 + 2l_ +   + 2l_ -   + k} dyd\phi  \\ 
 \end{array} }  \\ 
  = \frac{{\Gamma \left( {3 + l_ +   + l_ -   + \frac{k}{2}} \right)}}{2} \\ 
 \int {\frac{{\cos ^{^{2 + 2l_ +  } } \left( \phi  \right)\sin ^{^{2 + 2l_ -  } } \left( \phi  \right)}}{{\left( {\beta _ + ^2 \cos ^2 \left( \phi  \right) + \beta _ - ^2 \sin ^2 \left( \phi  \right)} \right)^{3 + l_ +   + l_ -   + \frac{k}{2}} }}d\phi }  \\ 
 \end{array}
\]
since the y integral is trivial. Then the k = 0 term is just the normalization: 
\[
\frac{1}{{A^2 }} = g_{l_ +  l_ -  }^0 \left( {\beta _ +  ,\beta _ -  } \right) = \frac{{\Gamma \left( {\frac{3}{2} + l_ +  } \right)\Gamma \left( {\frac{3}{2} + l_ -  } \right)}}{{4\beta _ + ^{3 + 2l_ +  } \beta _ - ^{3 + 2l_ -  } }}
\]
 so that 
\[
\begin{array}{l}
 f_{l_ +  l_ -  }^k \left( {\beta _ +  ,\beta _ -  } \right) = \frac{{g_{l_ +  l_ -  }^k \left( {\beta _ +  ,\beta _ -  } \right)}}{{A^2 }} \\ 
  = \frac{{2\Gamma \left( {3 + l_ +   + l_ -   + \frac{k}{2}} \right)\beta _ + ^{3 + 2l_ +  } \beta _ - ^{3 + 2l_ -  } }}{{\Gamma \left( {\frac{3}{2} + l_ +  } \right)\Gamma \left( {\frac{3}{2} + l_ -  } \right)}} \times  \\ 
 \int {\frac{{\cos ^{^{2 + 2l_ +  } } \left( \phi  \right)\sin ^{^{2 + 2l_ -  } } \left( \phi  \right)}}{{\left( {\beta _ + ^2 \cos ^2 \left( \phi  \right) + \beta _ - ^2 \sin ^2 \left( \phi  \right)} \right)^{3 + l_ +   + l_ -   + \frac{k}{2}} }}d\phi }  \\ 
 \end{array}
\]
This cannot be evaluated analytically for for k odd. However, checks that can be run include:
\[
f_{l_ +  l_ -  }^0 \left( {\beta _ +  ,\beta _ -  } \right) = 1,f_{l_ +  l_ -  }^2 \left( {\beta _ +  ,\beta _ -  } \right) = \frac{{\left( {l_ +   + \frac{3}{2}} \right)}}{{\beta _ + ^2 }} + \frac{{\left( {l_ -   + \frac{3}{2}} \right)}}{{\beta _ - ^2 }}
\]
so approximately 
\[
f_{l_ +  l_ -  }^1 \left( {\beta _ +  ,\beta _ -  } \right) = \left( {\frac{{\left( {l_ +   + \frac{3}{2}} \right)}}{{\beta _ + ^2 }} + \frac{{\left( {l_ -   + \frac{3}{2}} \right)}}{{\beta _ - ^2 }}} \right)^{1/2} 
\]

\section{n-body Interactions}
The trick in the previous section does not work for an $N_B$-bead solution with massive quarks, Instead we need to evaluate the expectation values of terms like 
\[
\left| {\hat r_1 } \right| = \sqrt {\hat r_1 ^2 }  = \sqrt {\left( {\sum\limits_{}^{} {e_j^{\left( 1 \right)} u_j } } \right)^2 }  = \sqrt {\sum\limits_{}^{} {e_j^{\left( 1 \right)} e_k^{\left( 1 \right)} u_j u_k } } 
\]
 We cannot evaluate 
$\int {\left| {\hat r_1 } \right|^k } e^{ - \sum\limits_{}^{} {\beta _i ^2 u_i^2 } } \Pi u_i^2 du_i$
for k = 1 directly.
Write 
\[
\begin{array}{l}
 F^N \left( {\beta _i } \right) = C\int {\left| {\sum\limits_{}^{} {e_i^2 u_i^2 } } \right|^{1/2} \prod\limits_i^N {e^{ - \beta _i ^2 u_i^2 } } } u_i^2 du_i^{}  \\ 
 E_n  = \sum\limits_{i = n + 1}^N {e_i^2 u_i^2 } ,B_n  = \frac{3}{2}\sum\limits_{i = 1}^n {\frac{{e_i^2 }}{{\beta _i ^2 }}}  \\ 
 \end{array}
\]
(C is the normalisation). Then we can expand, integrate and contract:
\[
\begin{array}{l}
 \int {\left| {\sum\limits_{}^{} {e_i^2 u_i^2 } } \right|^{1/2} e^{ - \beta _1 ^2 u_1^2 } u_1^2 du_1^{} }  = \int {\left| {A_1  + e_1^2 u_1^2 } \right|^{1/2} e^{ - \beta _1 ^2 u_1^2 } u_1^2 du_1^{} }  \\ 
  \approx \left( {A_1  + B_1 } \right)^{1/2}  = \left( {A_2  + B_1  + e_2^2 u_2^2 } \right)^{1/2}  \\ 
 \end{array}
\]
or in general 
\[
\int {\left| {A_n  + B_{n - 1}  + e_n^2 u_n^2 } \right|^{1/2} e^{ - \beta _n ^2 u_n^2 } u_n^2 du_n^{} }  \approx \left( {A_n  + B_n } \right)^{1/2} 
\]
Repeating this gives 
\[
F^N \left( {\beta _i } \right) \approx \sqrt {B_N }  = \left( {\frac{3}{2}\sum\limits_{i = n - 1}^N {\frac{{e_i^2 }}{{\beta _i ^2 }}} } \right)^{1/2} 
\]
. All the cross terms vanish on integration.

We can extend this: e.g. Coulomb term only exists between quarks, so write 
\[
\begin{array}{l}
 \frac{1}{{\left| {\vec r_1  - \vec r_N } \right|}} = \frac{1}{{\left| {\vec r_1  - \vec r_2  + \vec r_2  - \vec r_3  + ... - \vec r_N } \right|}} \\ 
  = \frac{1}{{\left| {\sum\limits_{}^{} {\hat r_i } } \right|}} = \left( {\sum\limits_{jl}^{} {\hat u_j \hat u_l } \sum\limits_{ik}^{} {e_k^l e_i^j } } \right)^{ - 1/2}  \\ 
 \end{array}
\]
giving the expectation value
\[
F_m ^C \left( {\beta _i } \right) \approx \left( {\frac{3}{2}\sum\limits_{i = n - 1}^N {\frac{{b_i^C }}{{\beta _i ^2 }}} } \right)^{1/2} ,b_i^C  = \sum\limits_{jk}^{} {e_j^i e_k^i } 
\]
\bibliography{Hybrid.bib}

\end{document}